\documentclass[aps,pra,twocolumn,preprintnumbers]{revtex4}
\usepackage{graphicx}
\usepackage{amstext}
\usepackage{array}
\usepackage{amssymb}
\usepackage{amsmath}
\usepackage{color}

\newcommand{\be}{\begin{equation}}
\newcommand{\ee}{\end{equation}}
\newcommand{\bea}{\begin{eqnarray}}
\newcommand{\eea}{\end{eqnarray}}
\newcommand{\Eq}[1]{Eq.\,(\ref{#1})}% \Eq{abc}
\newcommand{\Fig}[1]{Fig.\,\ref{#1}}% \Fig{fig:abc}
\newcommand{\Sec}[1]{Sec.\,\ref{#1}}% \Sec{sec:abc}
\newcommand{\Tab}[1]{Table\,\ref{#1}}% \Tab{tab:abc}
\newcommand{\Onlinecite}[1]{Ref.\,[\onlinecite{#1}]} %\Onlinecite{abc}

\newcommand{\GF}{\hat{\mathbf{G}}}

\newcommand{\br}{\mathbf{r}}

\newcommand{\bE}{\mathbf{E}}

\newcommand{\En}{\mathbf{E}_n}

\newcommand{\tEn}{\tilde{\mathbf{E}}_n}
\newcommand{\tEm}{\tilde{\mathbf{E}}_m}
\newcommand{\Em}{\mathbf{E}_{m}}
\newcommand{\Fn}{\mathbf{F}_n}

\newcommand{\heps}{\hat{{\pmb{\varepsilon}}}}

\newcommand{\hsigma}{\hat{\boldsymbol{\sigma}}}
\newcommand{\hs}{\hat{\boldsymbol{s}}}

\newcommand{\fe}{f_\mathrm{e}}
\newcommand{\fh}{f_\mathrm{h}}

\begin{document}
%\pagewiselinenumbers
\title{Applying the resonant-state expansion to realistic materials with frequency dispersion}
\author{H.\,S. Sehmi}
\author{W. Langbein}
\author{E.\,A. Muljarov}
\affiliation{School of Physics and Astronomy, Cardiff University, Cardiff CF24 3AA,
United Kingdom}
\begin{abstract}
The dispersive resonant-state expansion, developed for an accurate calculation of the resonant states in open optical systems with frequency dispersion, is
applied here to realistic materials, such as metallic nanoparticles and semiconductor microspheres. The material permittivity is determined by fitting the measured indices of refraction and absorption with a generalized Drude-Lorentz model containing a number of poles in the complex frequency plane. Each Drude or Lorentz pole generates an infinite series of resonant states. Furthermore, for small nanoparticles, each of these poles produces a distinct surface plasmon polariton mode. The evolution of these multiple surface modes with increasing radius traces the transition from the electrostatic limit to significant retardation and radiation. Treating the optical phonon range in a semiconductor microsphere, a reststrahlen band separating the resonant states is found.  Considering a small energy range around the semiconductor band gap, the transition from absorption to gain is described by inverting the Lorentz pole weight, which results in the formation of lasing resonant states.
Interestingly, the series of resonant states converging towards the absorption pole from the lower frequency side reshapes for a gain pole into a clockwise loop approaching the pole from the higher frequency side, being separated from a series spanning from low to high frequencies and containing the lasing modes.

\end{abstract}
%
%\pacs{03.50.De, 42.25.-p, 03.65.Nk}
%
\date{\today}
\maketitle

\section{Introduction}
Applying the concept of resonant states (RSs) to optical systems is a mathematically rigorous way of treating their electromagnetic resonances. RSs are the  eigenmodes of an optical system which can be found by solving Maxwell's wave equation with outgoing wave boundary conditions. In open systems, the RS eigenfrequencies $\omega_n$, numbered by the index $n$, are generally complex, which physically reflects the fact that the energy leaks out of the system. For an isolated resonance in an optical spectrum of the system, the real part of the corresponding $\omega_n$ determines the center frequency of the resonance and the imaginary part  its half width at half maximum, with the quality factor of the resonance given by half of the ratio of real to imaginary part.

Based on the concept of RSs, a rigorous perturbative method of calculation of the RSs of an arbitrary finite optical system (called perturbed system) has been developed~\cite{MuljarovEPL10}. This method, called the resonant-state expansion (RSE), converts the Maxwell wave equation into a linear matrix eigenvalue problem using the RSs of a simpler and usually analytically solvable system (called unperturbed system) as a basis for expansion.  In this way, the RSE is not limited to small perturbations and finds numerically exactly all the eigenmodes of the perturbed system. The RSE has been recently generalized~\cite{MuljarovPRB16} to optical systems with frequency dispersion, such as metal or semiconductor nanoparticles. This dispersive RSE (dRSE) is suited for materials with dispersion of the permittivity described by a generalized Drude-Lorentz model~\cite{SehmiPRB17}. In practice, one can find the parameters of this model -- generalized conductivities and complex poles of the permittivity --  by fitting the measured indices of refraction and absorption, as we have recently demonstrated for metals~\cite{SehmiPRB17} and semiconductors~\cite{SehmiPIERS17}.

The RSE can treat perturbations of arbitrary strength~\cite{MuljarovEPL10}, and can be superior to existing computational methods in electrodynamics, such as finite difference in time domain (FDTD) and finite element method (FEM), in terms of accuracy and efficiency~\cite{DoostPRA14,LobanovPRA17}. One of the main advantages of the RSE method is that it calculates the full spectrum of RSs in a wide frequency range striving towards completeness, and it does not depend on further approximations, i.e. it is numerically exact.
Indeed, the only parameter of the RSE is the number of the RSs included in the basis, which can in principle be made arbitrarily large. While the formalism of the RSE is analytically advanced, its technical implementation is rather straightforward, as it reduces solving Maxwell's wave equation to a diagonalization of a matrix containing a perturbation. For the latter, one needs a  complete set of RSs of an unperturbed system, ideally chosen in such a way that the matrix elements of the perturbation can be efficiently computed.

In the present work we apply the dRSE to metal and semiconductor nanoparticles, using as basis system a non-dispersive dielectric sphere. This approach has been used in \Onlinecite{MuljarovPRB16} for a dispersion described by the Drude model. Here, we extend the method to systems with a dispersion of the permittivity described by up to four pairs of Lorentz poles. Including these poles in the dispersion leads to new physical insights into the properties of metal and semiconductor particles, such as a coexistence of multiple surface plasmon polariton modes, which is of significance for the field of plasmonics \cite{ShahbazyanBook14}. We also explore how the dRSE can describe the onset of lasing when material absorption changes to gain in semiconductor microstructures, important for the field of optical microcavities \cite{VahalaBook04,KavokinBook07} and semiconductor lasers \cite{ChowBook12}.

The paper is organized as follows. In \Sec{RSE} we outline the formalism of the dRSE, providing general expressions for the normalization of dispersive RSs and for linear matrix eigenvalue problem determining the perturbed resonant states. The dRSE formalism is based on the condition that the dispersion of the unperturbed and perturbed systems has the same pole structure. In order to allow the unperturbed system to have no dispersion, or the perturbation to introduce additional poles in the dispersion, we present in \Sec{epsDisp} a variation of the dRSE called infinitesimal dispersive RSE (idRSE), which extends the basis of the unperturbed RSs to include a subset of RSs, called pole RSs, created by the additional poles of the permittivity in the limit of infinitesimal conductivities. We then apply in \Sec{sec:results} the dRSE to metal and semiconductor spherically symmetric systems. Using the idRSE, we find in \Sec{metals} the RSs of a gold sphere described by the Drude-Lorentz model with 3 pairs of Lorentz poles. This is done using the RSs of a dielectric sphere as basis extended to include the pole RSs. We concentrate our study on the surface plasmon polariton RSs and investigate their dependence on the sphere radius changing from 1 to 100\,nm. In \Sec{GaAs} we consider the optical-phonon range of the permittivity of GaAs, which can be accurately fitted by 1 pair of Lorentz poles, and find the RSs of a 50\,$\mu$m-radius GaAs microsphere. We study a small shift of these Lorentz poles in \Sec{1stOrder} by using a diagonal approximation of the dRSE which is corresponding to a 1st-order perturbation theory. Finally, in \Sec{Gain} we calculate the lasing modes of GaAs microspheres with 10\,$\mu$m and 1\,$\mu$m radius, having the permittivity fitted close to the GaAs band edge by 4 pairs of classical Lorentz poles. The optical gain is introduced by inverting the weight of the pole pair at the band edge from absorption to gain.  Details of the permittivity models and the calculation of basis RSs and their matrix elements are given in Appendices~\ref{App:DL}--\ref{App:basis}.

\section{Formalism of the dispersive RSE}
\label{RSE}

The basis for the dRSE is formed from the RSs of an unperturbed system which are the eigensolutions of the Maxwell wave equation
\be
\label{eqn:ME}
\nabla\times\nabla\times\En(\br)=\frac{\omega_n^2}{c^2}\,\heps(\br,\omega_n)\En(\br)
\ee
with outgoing wave boundary conditions (here $c$ is the speed of light in vacuum). The permittivity tensor $\heps(\br,\omega)$ of the unperturbed system is taken as analytic function in the complex frequency plane, with a countable number of simple poles. According to the Mittag-Leffler theorem~\cite{ArfkenBook13}, the frequency dispersion of the permittivity tensor can be expressed as
\be
\heps(\br,\omega)=\heps_\infty(\br)+\sum_j\frac{i\hsigma_j(\br)}{\omega-\Omega_j}\,,
\label{eqn:eps}
\ee
where $\heps_\infty(\br)$ is the high-frequency value of the permittivity and $\Omega_j$ are the resonance frequencies, at which the permittivity has simple poles, with the residues at the poles given by generalized conductivity tensors $\hsigma_j(\br)$. Equation~(\ref{eqn:eps}) is known as generalized Drude-Lorentz model~\cite{SehmiPRB17}; more details about the model and its special cases (Drude and Lorentz models) are provided in Appendix~\ref{App:DL}.

%To apply the RSE to 3D systems, we need a complete basis of RSs. We choose here the analytically known RSs of a dielectric sphere of radius $R$  and permittivity $\varepsilon(\omega)$, surrounded by vacuum. For any spherically symmetric system, the solutions of Maxwell's equations split into four groups: TE, TM, longitudinal electric (LE), and longitudinal magnetic (LM) modes~\cite{StrattonBook41}. In this paper we will be considering TM solutions only, which are encompassing surface plasmons.

The dRSE solves the Maxwell wave equation
\be
\label{eqn:ME1}
\nabla\times\nabla\times\bE(\br)=\frac{\omega^2}{c^2}\,\left[\heps(\br,\omega)+ \Delta\heps(\br,\omega)\right]\bE(\br)
\ee
for the perturbed system (the system of interest) described by a modified permittivity tensor $\heps(\br,\omega)+ \Delta\heps(\br,\omega)$, with the perturbation
\be
\Delta\heps(\br,\omega)=\Delta\heps_\infty(\br)+\sum_j\frac{i\Delta\hsigma_j(\br)}{\omega-\Omega_j}
\label{eqn:eps1}
\ee
having the {\em same} pole structure as $\heps(\br,\omega)$, i.e. with the poles at the same frequencies $\Omega_j$ and frequency-independent tensors $\Delta\heps_\infty(\br)$ and $\Delta\hsigma_j(\br)$. The electric field $\bE(\br)$ and the frequency $\omega$ in \Eq{eqn:ME1} are, respectively, the wave function and the eigenfrequency of a RS of the perturbed system. The perturbed Maxwell's wave equation (\ref{eqn:ME1}) can be solved with the help of the Green's function (GF) $\GF_\omega(\br,\br')$ of the unperturbed \Eq{eqn:ME} in the following way:
\be
\bE(\br)=-\frac{\omega^2}{c^2} \int \GF_\omega(\br,\br') \Delta\heps(\br',\omega) \bE(\br') d\br'\,.
\label{eqn:E1}
\ee

The completeness of the basis of the unperturbed RSs allows us to convert this integral equation to a matrix eigenvalue problem.
This can be done by expanding the perturbed RS field into the unperturbed ones,
\be
\bE(\br) = \sum_n c_n\En(\br)\,,
\ee
and by using the Mittag-Leffler theorem again, this time expanding the Green's dyadic,
\be
\GF_\omega(\br,\br')=c^2\sum _n\frac{\En(\br)\otimes\En(\br')}{\omega_n(\omega-\omega_n)}\,.
\label{eqn:ML}
\ee
Here $\otimes$ denotes the dyadic product of the vector fields, and the RS wave functions $\En(\br)$ are properly normalized~\cite{DoostPRA14,MuljarovPRB16,MuljarovPRB16a,MuljarovPRA17}.
For non-static RSs, i.e. those having $\omega_n\neq 0$, the correct normalization linked to \Eq{eqn:ML} is given by~\cite{MuljarovPRB16,MuljarovPRB16a,MuljarovOL18}
\bea
1&=&2\int_{V}\En (\br)\cdot\left.\frac{\partial\bigl(\omega^2\heps(\br,\omega)\bigr)}{\partial(\omega^2)}\right|_{\omega_n}\En(\br)\,d{\bf r}
\nonumber
\\
&&+\frac{c^2}{\omega^2_n}\oint_{S_V} \left(\En\cdot\frac{\partial\Fn}{\partial
s}-\Fn \cdot \frac{\partial \En}{\partial
s}\right) dS\,,
 \label{norm}
\eea
where $\Fn=(\br\cdot\nabla)\En$, $V$ is an arbitrary volume containing the system, surface $S_V$ is its boundary, and
$\partial/\partial s$ is the spatial derivative  along the outer surface normal. Note that the electric field normalized according to \Eq{norm} is a factor $\sqrt{2}$ smaller than the one used in the previous works~\cite{MuljarovEPL10,MuljarovPRB16}. This has been introduced in order to symmetrize the normalization with respect to the electric and magnetic field (for more details see \Onlinecite{MuljarovOL18}).

The sum in \Eq{eqn:ML} is taken over all RSs; their eigenfrequencies $\omega_n$ play the role of simple poles of the GF. Owing to the frequency dispersion \Eq{eqn:eps}, the GF has alternative spectral representations linked to \Eq{eqn:ML} via sum rules which the RSs of the dispersive system obey~\cite{MuljarovPRB16,MuljarovOL18}. Using in \Eq{eqn:E1} a proper combination of the different spectral representations of the GF and equating the coefficients at the basis functions $\En(\br)$, this integral equation  reduces to the following matrix eigenvalue problem, linear in frequency $\omega$:
\be
\omega_n\sum _m\left(\delta_{nm}-U_{nm}\right)c_{m}=\omega\sum _m\left(\delta_{nm}+V_{nm}\right)c_{m}\,,
\label{eqn:RSE}
\ee
which presents the key equation of the dRSE. Here, $V_{nm}$ and and $U_{nm}$ are the matrix elements of the non-dispersive and dispersive parts of the perturbation, given by
\be
V_{nm} = \int \En(\br)\cdot\Delta\heps_\infty(\br)\Em(\br)\,d \br
\ee
and
\begin{align}
U_{nm} &=\int \En(\br)\cdot[\Delta\heps(\br, \omega_n)-\Delta\heps_\infty(\br)]\Em(\br)\,d \br
\label{eqn:unm1}
\\
&=\sum\limits_j \frac{i}{\omega_n-\Omega_j}\int \En(\br)\cdot\Delta\hsigma_j(\br)\Em(\br)\,d \br\,,
\label{eqn:unm}
\end{align}
respectively. For a detailed derivation of the dRSE equation (\ref{eqn:RSE}), see Refs.~\cite{MuljarovPRB16,MuljarovOL18}.

\section{Infinitesimal-dispersive basis}
\label{epsDisp}

The dRSE formally requires the permittivity of the basis system and the perturbed system to have the same poles. While this is looking like a limitation of the dRSE, in reality, the perturbation $\Delta\heps(\br, \omega)$ can introduce new poles in the permittivity. An extreme case of this situation (illustrated in \Sec{sec:results}) is having an unperturbed system without frequency dispersion, such as a dielectric sphere in vacuum, but its RSs are used as a basis for the dRSE treating a system with dispersion, such as a metal or a semiconductor nanoparticle. However, in the functional space of dispersive solutions, the basis of non-dispersive RSs is incomplete and needs to be complemented (for completeness) with so-called {\it pole} RSs (pRSs).

The pRSs originate from a group of RSs of a dispersive system with a permittivity pole at $\omega=\Omega_j$ having an infinitesimal conductivity, $\hsigma_j\to0$. All the pRSs due to this pole then become degenerate, $\omega_n\to\Omega_j$, while having different values of the effective refractive index, and hence different wave functions. These wave functions, however, have vanishing normalization constants, $A_n\to0$, so that without perturbing the pole to a finite conductivity, the pRSs do not contribute to the modes of the system, as both $V_{nm}$ and $U_{nm}$ are vanishing for $n$ or $m$ referring to pRS.

To treat the pRSs of the $j$-th pole of the permittivity, we introduce $i\hsigma_j(\br)=\xi \hs_j(\br)$ with $\xi\to 0$. $\hs_j(\br)$ is an arbitrary finite tensor, which can be
chosen, without loss of generality, as $\hs_j(\br)=\hat{\mathbf{1}}$ within the volume $V$ of the system and zero outside it. In the limit $\xi\to 0$, the eigenfrequencies of the pRSs
%due to this pole of the permittivity
become
\be
\omega_n=\Omega_j+\xi q_{n}\,,
\label{poleRS}
\ee
where $q_{n}$ are finite eigenvalues determined by the geometry of the system and its material parameters.
For example, for transverse-magnetic (TM) modes of a sphere in vacuum they are found from solving a secular equation (\ref{secular}), see Appendix~\ref{App:TMmodes}.
Note that for brevity of notations, here and below we have omitted index $j$ which would appear in all the eigenvalues and matrix elements due to the pRSs of pole $j$.

According to Eqs.\,(\ref{eqn:eps}) and (\ref{poleRS}), the permittivity calculated at the pRS frequency $\omega_n$, contributing both to Eqs.(\ref{norm}) and (\ref{eqn:unm1}), is given by
\be
\heps(\br, \omega_n)=\heps_\infty(\br)+\sum_{j'\neq j}\frac{i\hsigma_{j'}(\br)}{\Omega_j-\Omega_{j'}} +\frac{\hs_j(\br)}{q_{n}}\,,
\label{qn}
\ee
which is a finite quantity, determining the effective refractive index and hence the wavelength of the pRS. However, the frequency derivative of the permittivity, contributing to the RS normalization \Eq{norm}, is divergent,
\be
\left.\frac{\partial\heps(\br,\omega)}{\partial(\omega)}\right|_{\omega_n} =-\frac{1}{\xi}\frac{\hs_j(\br)}{q_{n}^2}\,,
\label{eps-der}
\ee
which leads to a vanishing normalization coefficient of the pRS field, as follows from  \Eq{norm}:
\be
\xi=-\frac{\Omega_j}{q_{n}^2}\int_{V}\En(\br)\cdot\hs_j(\br)\En(\br)\,d{\bf r}\,,
\label{polenorm}
\ee
since $\xi\to 0$ but ${\Omega_j}/{q_{n}^2}$ is finite.

For any finite perturbation of the conductivity $\Delta\hsigma_j\neq0$, the vanishing normalization constants of pole RSs are compensated in the perturbation matrix by divergencies due to the dispersion, $i\Delta\hsigma_j/(\omega_n-\Omega_j)\propto 1/\xi$, thus making $U_{nm}$ finite. To take this limit properly, it is convenient to introduce the factors
\be
\alpha_n= \left\{
\label{eqn:alpha}
\begin{array}{cl}
	1 & \text{for non-pRSs}\\
&\\
	\displaystyle \sqrt{\frac{\omega_n-\Omega_j}{\Omega_j}} & \text{for pRSs of $\Omega_j$}\\
\end{array} \right.
\ee
and to re-define the RS wave functions as
\be
\En(\br)=\alpha_n\tEn(\br)\,,
\label{Ent}
\ee
where all $\tEn(\br)$ have finite normalization.
Expressing the perturbation matrices and eigenvectors as
\begin{align}
V_{nm} &= \alpha_n\alpha_m Q_{nm}, & U_{nm} &=\frac{{\alpha_m}}{\alpha_n}S_{nm}, & c_n &= \frac{b_n}{\alpha_n},
\end{align}
the dRSE matrix equation (\ref{eqn:RSE}) takes the form:
\be
\omega_n\sum _m\left(\delta_{nm}-S_{nm}\right)b_{m}=\omega\sum _m\left(\delta_{nm}+\alpha_n^2 Q_{nm}\right)b_{m}\,,
\label{eqn:epsDispRSE}
\ee
in which the elements $Q_{nm}$ and $S_{nm}$ are finite. Explicitly,
the matrix $Q_{nm}$ is defined according to
\be
Q_{nm} = \int \tEn(\br)\cdot\Delta\heps_\infty(\br)\tEm(\br)\,d \br\,,
\ee
while the matrix $S_{nm}$ is given by
\be
S_{nm} = \int \tEn(\br)\cdot[\Delta\heps(\br, \omega_n)-\Delta\heps_\infty(\br)]\tEm(\br)\,d \br\,,
\ee
if $n$ refers to a non-pRS, and by
\be
S_{nm} =\frac{i}{\Omega_j} \int \tEn(\br)\cdot\Delta\hsigma_j(\br)\tEm(\br)\,d \br\,,
\label{S-pole}
\ee
if $n$ refers to a pRS with $\omega_n=\Omega_j$. Accordingly, the perturbed field is given by
\be
\bE(\br) = \sum_n b_n\tEn(\br)
\ee
in terms of the modified basis RS wave functions $\tEn(\br)$ and the eigenvector components $b_n$ determined by \Eq{eqn:epsDispRSE}. Clearly, if $\hsigma_j(\br)=\Delta\hsigma_j(\br)=0$, none of the corresponding pRSs is expected to contribute to the perturbed field. Indeed, in this case $S_{nm}=0$ for any $j$-th pRS $n$, according to \Eq{S-pole}, while $\alpha_n=0$ is leading to $\alpha_n^2Q_{nm} =0$ in \Eq{eqn:epsDispRSE}, even though $V_{nm}\neq0$. Hence $b_n=0$, showing that this pRS does not contribute to the perturbed RSs.

The method presented in this section is called {\em infinitesimal-dispersive} RSE (idRSE). The described procedure is suited for any number of new poles introduced in the perturbation. It has been already applied to systems with dispersion described by the Drude model~\cite{MuljarovPRB16}. In \Sec{sec:results} below we illustrate the idRSE on several examples, for the dispersion of the permittivity including both Drude and Lorentz poles, or Lorentz poles only.

\section{Results}
\label{sec:results}

In this section we present a selection of applications of the dRSE. To provide a clearer illustration of the physical effects treated, and to compare results of the dRSE with available exact solutions, we concentrate here on spherically symmetric systems, but note that the RSE can also be applied to non-spherical systems~\cite{DoostPRA14}.
We consider TM polarization, and treat $l=1$ modes of metallic particles, showing localized surface plasmons (SP), where $l$ is the orbital quantum number. For semiconductor microspheres instead, we use $l\gg1$ in order to obtain sharp whispering-gallery (WG) modes. The wave functions, normalization coefficients, and the secular equation determining the RS eigenvalues of a uniform sphere in vacuum
%the frequency eigenvalues of the normal RSs and the refractive-index eigenvalues of pole RSs
are provided in Appendix~\ref{App:TMmodes}, along with the matrix elements for uniform perturbation.
Approximate recursive relations determining the start values for solving the secular equation are given in Appendix~\ref{App:basis}.
The basis for the dRSE is truncated in such a way that all unperturbed RSs satisfying the condition $|n_r(\omega_n)\omega_n|<\Omega_{\rm c}$ are included in the basis, where $n_r(\omega)=\sqrt{\varepsilon(\omega)}=n_r'+i n_r''$ is the complex refractive index, and $\Omega_{\rm c}$ is the cut-off frequency determining the basis size $N$. For each $l$, a longitudinal static mode~\cite{DoostPRA14} with $\omega_n=0$ has been included in the basis; this is required for completeness of a non-dispersive basis or a basis without Ohm's law pole in the dispersion.

To apply the dRSE to spherical particles made of realistic materials, such as gold and GaAs considered below, we have generated the corresponding parameters of the Drude-Lorentz model using the earlier developed fit program~\cite{SehmiPRB17}. This fit program is using the experimentally measured indices of refraction and absorption, $n_r'$ and $n_r''$, respectively, in order to find the optimal values of the pole frequencies $\Omega_j$ and their associated conductivities $\sigma_j$. The value of the high-frequency permittivity $\varepsilon_\infty$ can be used as an adjustable parameter which can be chosen to generate a set of the basis RSs providing quick convergence with $N$. We have found in particular that the value of $\varepsilon_\infty=1$ used for the basis system corresponds to the slowest convergence as in this case the refractive-index contrast between the system and the environment (vacuum) is minimized for high-frequency modes. In the example provided in \Sec{metals} we transform a dielectric (silica) sphere with $\varepsilon_\infty=2.1272$ to a gold sphere with $\varepsilon_\infty+\Delta\varepsilon_\infty=0.5$, while in Secs.\,\ref{GaAs}--\ref{Gain} we use $\varepsilon_\infty=11.0$ or 8.6013 for calculating the RSs of a GaAs microsphere,  while keeping $\Delta\varepsilon_\infty=0$. \Tab{tab:para4} of Appendix~\ref{App:DL} shows all the fit parameters of gold and GaAs used in the dRSE calculations presented in this section below.

\subsection{Infinitesimal-dispersive RSE: Plasmons}
\label{metals}

\begin{figure}
\includegraphics[width=\columnwidth]{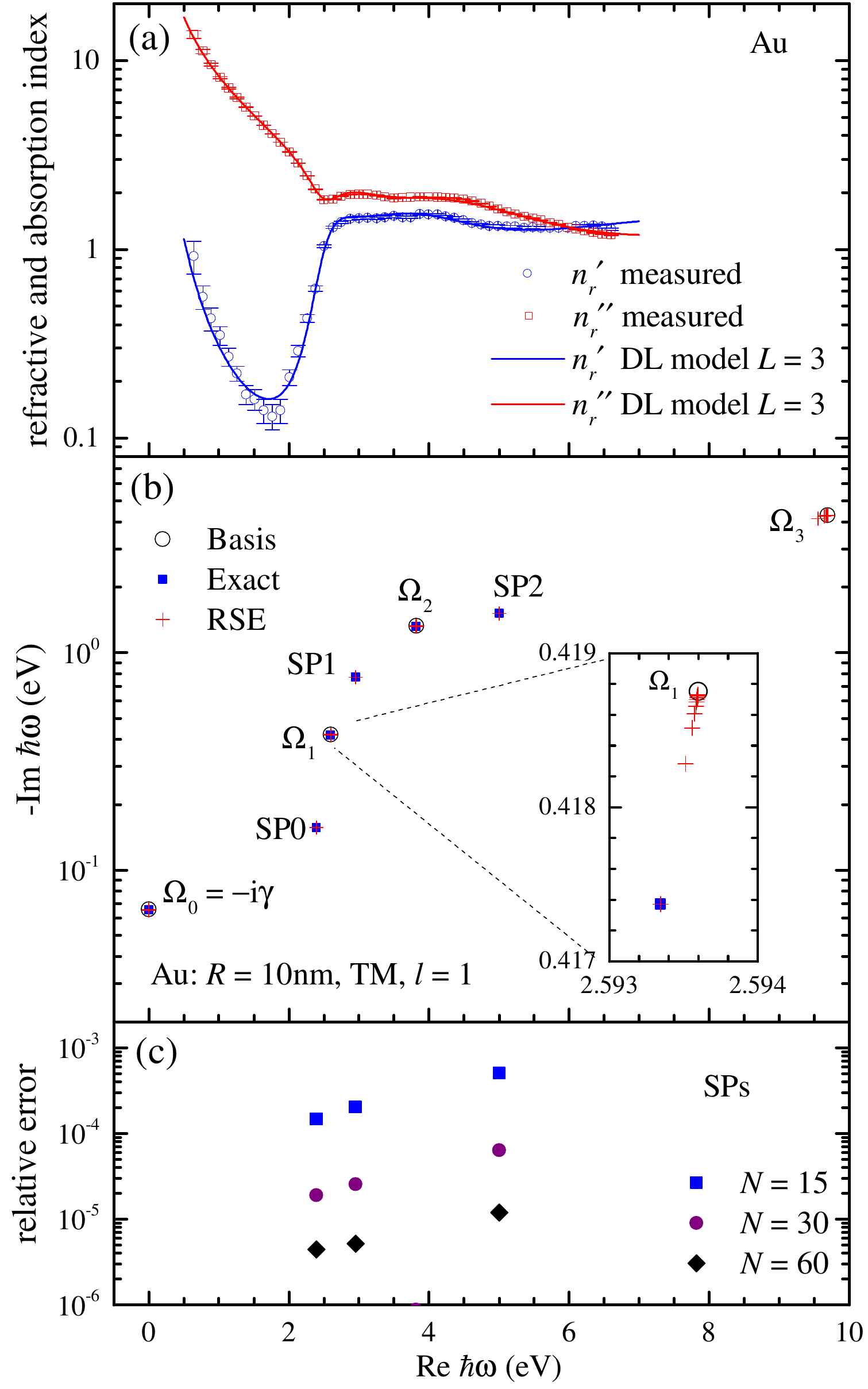}
\caption{(a) Refractive index $n_r'$ and absorption index $n_r''$ of gold according to \Onlinecite{JohnsonPRB72} (circles and error bars) and their fit using the Drude-Lorentz model \Eq{eqn:DL} with $L=3$ (solid lines) versus photon energy $\hbar\omega$. (b) RSs for sand ($n=1.4585$) to gold (see \Tab{tab:para4}) idRSE for a nanosphere of radius $R=10$\,nm in vacuum, for $l=1$ and TM polarization. The basis RSs in the frequency region shown are only the pole RSs. The exact solutions are found by solving the secular equation (\ref{secular}). The inset shows the first Lorentz pole-related RSs. (c) Relative error of idRSE for the three surface plasmon modes versus basis size $N$, showing an error scaling as $N^{-3}$.}
\label{fig:gold10nm}
\end{figure}

To construct a basis for the idRSE treating a gold nanosphere, we add to the set of the RSs of a non-dispersive dielectric nanosphere with $n_r=1.4585$ the pRSs of a Drude pole $\Omega_0=-i\gamma$ and of 3 pairs of Lorentz poles $\Omega_j$  ($j=\pm1$, $\pm2$ and $\pm3$) with infinitesimal conductivities. The poles are shown in \Fig{fig:gold10nm}(b), lined up with the corresponding fit of $n_r'$ and $n_r''$ of gold~\cite{JohnsonPRB72}, which is illustrated in \Fig{fig:gold10nm}(a).

The spectrum of the perturbed system -- a gold sphere of radius $R=10$\,nm -- shown in \Fig{fig:gold10nm}(b) reveals RSs to the high energy side of each pole in the dispersion. These are known as surface plasmon (SP) polariton resonances.
In the electrostatic limit, they occur at $\varepsilon(\omega)=-2$ for $l=1$, as given by \Eq{equ:SP} of Appendix~\ref{App:TMmodes}, and are well known in the literature \cite{BohrenBook98}. This equation provides one SP state for each pole in $\varepsilon$. Accordingly, using the Drude model, only one such RS is present~\cite{MuljarovPRB16}. Notably, the RSE calculates these poles beyond the electrostatic limit -- the resulting evolution of the SPs versus radius is discussed below. The validity of the response predicted by the SPs at complex frequencies depends on the reliability of the DL model of $\varepsilon(\omega)$ at these frequencies, keeping in mind that the parameters of the model are fitted to describe the permittivity at real frequencies. At small imaginary parts of the SP frequencies, such as for SP0, the validity is expected to be better than at larger imaginary parts, such as for SP1 and SP2. Importantly, for correct calculation of the response of the system at real frequencies, such as the scattering and absorption spectra, the complete set of its RSs is required~\cite{LobanovPRA18}, which necessarily includes all the SP modes generated by the model of $\varepsilon(\omega)$ used.
The absorption lineshape is therefore not just a single resonance, but is more complex, as also observed in the experiment.

In addition to the SP mode around each pole of the permittivity, the full solution of the secular equation (see \Eq{secular} of Appendix~\ref{App:TMmodes}) contains a countable infinite number of modes in the vicinity of the Drude and each Lorentz pole, originating from the pRSs. The inset in \Fig{fig:gold10nm}(b) shows these RSs approaching the first Lorentz pole $\Omega_1$. We can also see from the figure that the idRSE matches the exact perturbed solution; however, in the presence of Lorentz poles, it is difficult to find all modes within a frequency range reliably from the secular equation (\ref{secular}). The idRSE can find these solutions due to the completeness of RSs and is expected to work equally well for systems without known analytic solutions~\cite{DoostPRA14}. The relative error of the idRSE, given in \Fig{fig:gold10nm}(c) for the SPs,  shows a $N^{-3}$ dependence on the number of states in the basis, as observed previously for spherically symmetric systems~\cite{MuljarovEPL10,MuljarovPRB16}.

\begin{figure}
\includegraphics[width=\columnwidth]{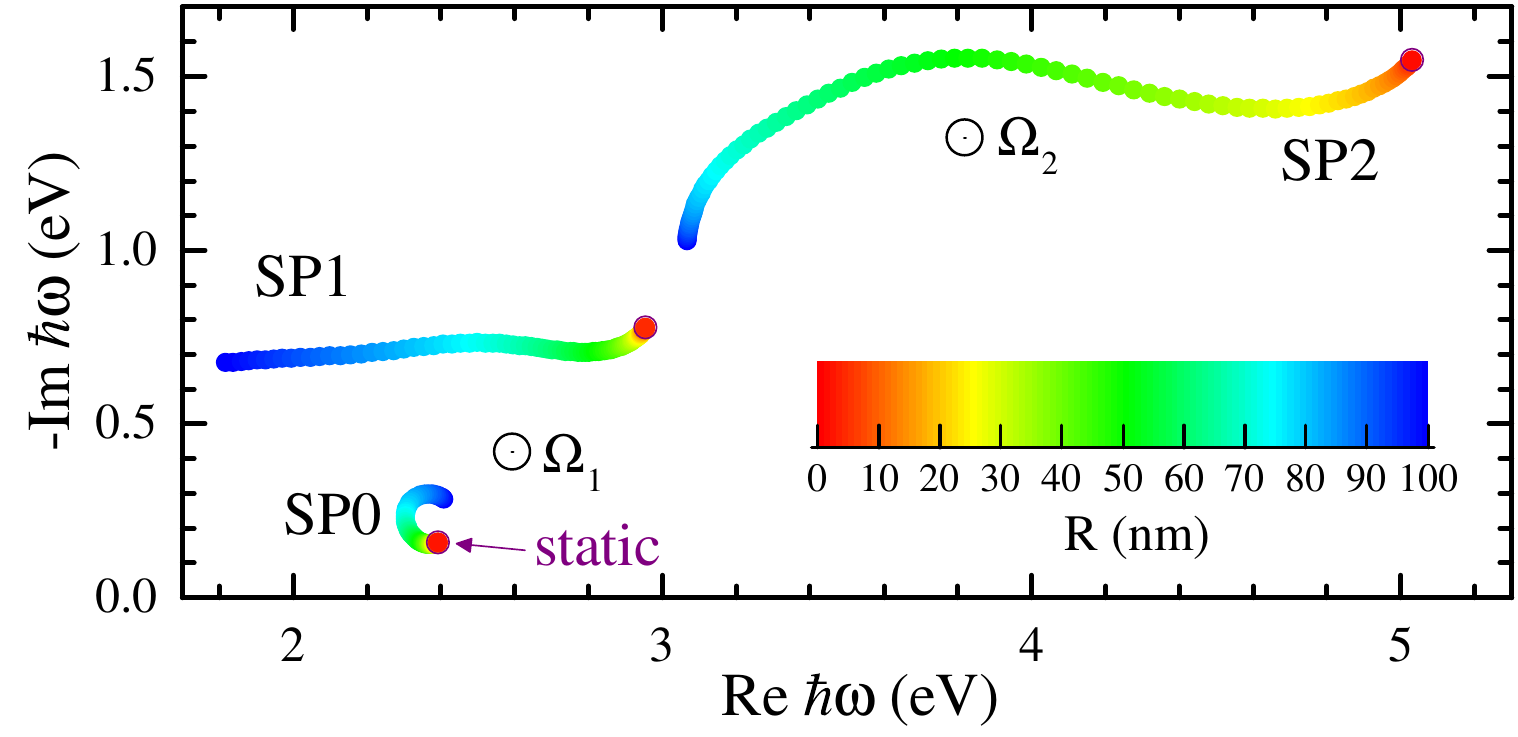}
\caption{Real and imaginary part of the energy of the three lowest SPs with $l=1$ (dipolar modes, TM polarization), with increasing radius $R$ of a gold nanosphere from 1\,nm to 100\,nm as given by the color code. The first and second Lorentz pole frequencies $\Omega_{1,2}$ are shown for reference. The static solutions ($\varepsilon(\omega)=-2$) are given as small black circles, and are consistent with the small $R$ limit of the SPs. }
\label{fig:surfacePlasmon}
\end{figure}

The evolution of the SP modes with nanosphere radius $R$ changing from 1\,nm to 100\,nm is given in \Fig{fig:surfacePlasmon}.
We find that for $R<15$\,nm, the SP modes keep close to the electrostatic limit, as expected for sizes much smaller than the light wavelength, $R\omega/c\ll1$, noting that the SP wavelengths are in the 250 to 500\,nm range. Increasing $R$, different SP modes change differently. Generally, they shift to longer wavelength, as expected from modes determined by size rather than dispersion. SP0, related to the Drude pole and thus representing the classical SP of gold, is not changing much in the full range of the radii shown, which is due to the dominant role of the Drude part in the dispersion of gold and a rather small value of $\gamma$ for the Drude pole, see \Tab{tab:para4}. For small $R$, SP0 has an energy around 2.4\,eV and a linewidth of about 0.3\,eV, comparable to the  features seen in the measured absorption \cite{BohrenBook98}. Increasing $R$, SP0 moves along a small arc, first shifting to the red, then increasing the magnitude of the imaginary part of the wave number due to the increasing role of the radiative broadening, and then shifting back to the blue, possibly due to a repulsion from SP1, the next closest surface plasmon mode originating from the Lorentz pole at $\Omega_1$.  The next SP mode, SP1, has an energy of 2.9\,eV in the static regime and a linewidth of 1.5\,eV; this mostly accounts for the experimentally observed intraband absorption features.  Note however that in calculating the absorption and scattering properties, the SPs will interfere \cite{LobanovPRA18}, so that the lineshape is expected to be non-Lorentzian, again in agreement with experiment. Calculating these properties using the dRSE will be the topic of a future work. With increasing $R$, SP1 is approximately keeping its linewidth, but is red-shifting, passing SP0 at around 80\,nm. This leads to a redshift and broadening of the observed SP resonance in absorption, in qualitative agreement with experiment \cite{MyroshnychenkoCSR08}. SP2 shows a similar behavior as SP1, but with a twice broader linewidth, therefore describing a broad absorption in the interband absorption range of gold.

\subsection{Infinitesimal-dispersive RSE: Optical phonons}
\label{GaAs}

The optical phonons in GaAs provide a sharp resonance in the permittivity over the energy range of $28-40$\,meV. This is well described by a single Lorentz pole with a purely imaginary positive conductivity, corresponding to a classical Lorentz oscillator having a purely real negative weight of the permittivity pole. The refractive and absorption index spectra and their fit are shown in \Fig{fig:GaAs50micron}(a). The fit parameters are given in \Tab{tab:para4} of Appendix~\ref{App:DL}. We note that from about 33 to 36\,meV, the absorption index supersedes the refractive index, which is the so-called reststrahlen region where a negative real part of the permittivity is causing an increased reflectivity of the material.

\begin{figure}
\includegraphics[width=\columnwidth]{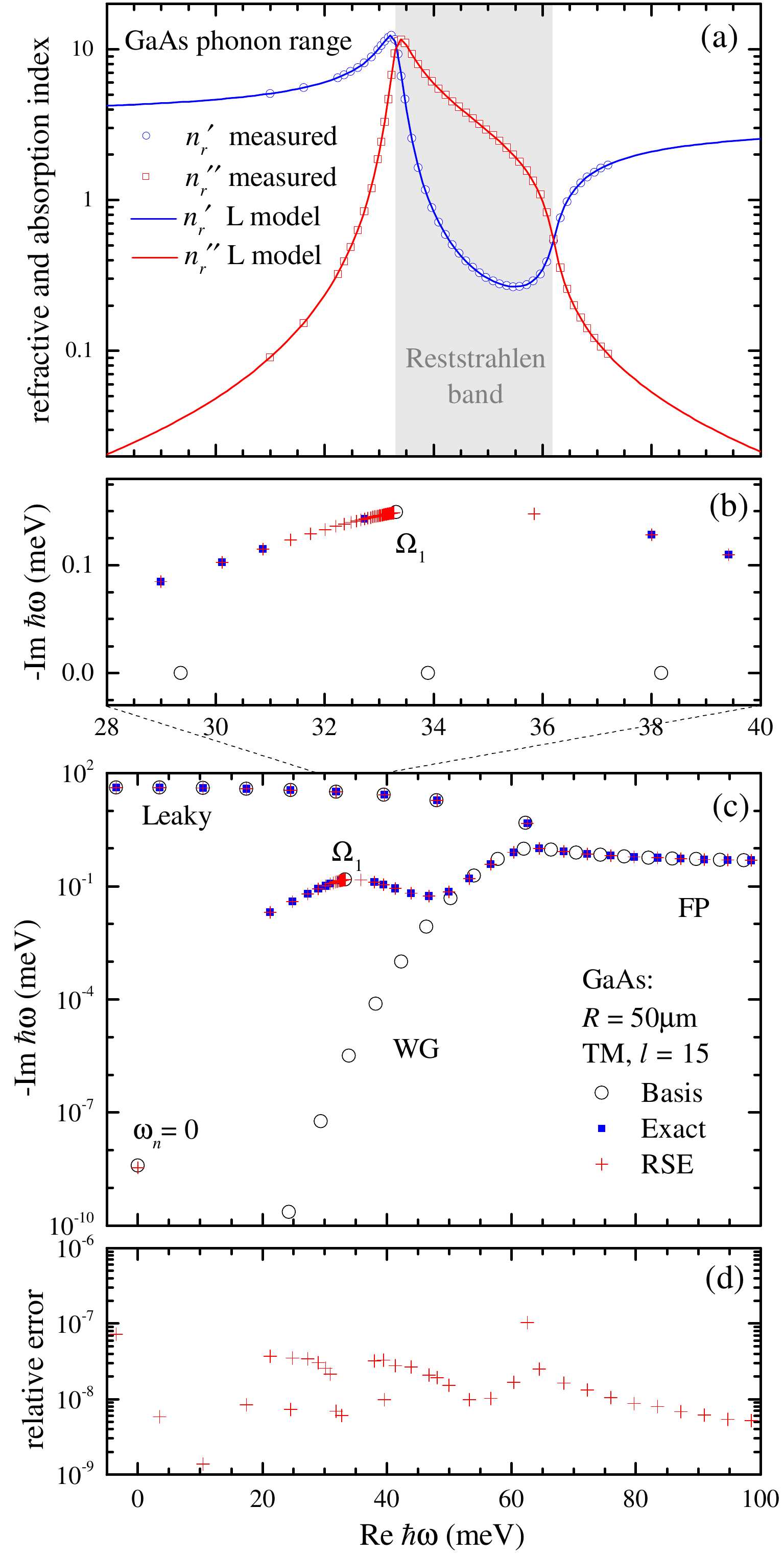}
	\caption{(a) Refractive index $n_r'$ and absorption index $n_r''$ of GaAs in the optical phonon range \cite{BlakemoreJAP82} (circles) and the Lorentz model \Eq{eqn:L} with $L=1$ (solid lines) as functions of the photon energy $\hbar\omega$. (b) RSs for $l=15$ (TM polarization) of the basis sphere of radius $R=50\,\mu$m with $n_r=3.317$ (open circles) and of the perturbed sphere with the GaAs optical phonon resonance perturbation showing the Lorentz pole ($\Omega_1$) related RSs, and the reststrahlen band around the resonance where there are no RSs. idRSE results for $N=319$  are shown as red crosses, and exact RSs using \Eq{secular} are shown as blue squares. (c) RSs over a larger frequency range, showing also the static RS with $\omega_n=0$. (d) relative error of RS frequencies calculated by the RSE. }
	\label{fig:GaAs50micron}
\end{figure}

We use the idRSE with a basis system given by a dielectric microsphere with the same $\varepsilon_\infty$ as used in the above fit. We choose $l=15$ and $R=50\,\mu$m  in order for the Lorentz pole $\Omega_1$ to be close to WG mode frequencies.
The RS frequencies of the basis system and of the perturbed system calculated via idRSE and exactly via \Eq{secular}, are shown in \Fig{fig:GaAs50micron}(c). Due to the large $l$, the basis is clearly separated into WG, Fabry-Perot (FP), and leaky RSs. In the new system, we find the WG RSs strongly modified. There is a series of RSs approaching $\Omega_1$ from the lower frequency side, see also a zoom in \Fig{fig:GaAs50micron}(b). This series is countable infinite, limited in the numerical calculation by the finite basis size used. It resembles the series of basis RSs with increasing frequency and accordingly reducing wavelength. In fact, approaching from the lower frequency side the pole at  $\Omega_1$, which is also the transversal optical phonon frequency, the real part of the permittivity increases without bound, and so does the refractive index $n_r'$, in this way reducing the light wavelength within the sphere towards zero.  The imaginary part of the perturbed RSs is, however, different from that of the WG modes, as it is dominated by the material absorption of the perturbed system. On the right side of $\Omega_1$ instead, the refractive index is reduced when approaching the longitudinal optical phonon frequency, separated from  $\Omega_1$ by the reststrahlen band. Consequently, the light wavelength within the sphere is increased. This produces a set of RSs which can be labeled with a radial quantum number reducing  towards the frequency at which the refractive index is zero. The leaky RSs are, in turn, nearly unaffected by the perturbation within the sphere, which is understandable considering their little spatial overlap with the sphere.

We again note that it is difficult to find all the exact solutions corresponding to the series of RSs near the pole of the permittivity, which is seen as the absence of some of the exact RS frequencies in \Fig{fig:GaAs50micron}(b). However, we found that using the RS frequencies determined by the idRSE as starting points for solving numerically the secular equation (via e.g. a gradient decent method) enabled us to overcome this difficulty. We also note that it is important for the accuracy and convergence of the dRSE to the exact result with increasing basis size, that no RSs are missed in the basis, which is selected using the high spatial frequency cutoff condition $|n_r(\omega_n)\omega_n|<\Omega_{\rm c}$. The relative error for $N=319$, shown in \Fig{fig:GaAs50micron}(c), is in the $10^{-8}$ range. This error is significantly smaller than the ones observed for perturbations of $\varepsilon_\infty$ for a similar basis size \cite{DoostPRA14, MuljarovEPL10}, a finding which we attribute the perturbation being localized in frequency. Once the frequency cutoff of the basis is much larger than the frequency of the perturbed pole (here $\Omega_{\rm c}\gg\Omega_1$), the effect of the perturbation on the RSs not included in the basis is small, as suggested by the denominator $\omega_n-\Omega_j$ in the corresponding matrix elements given in \Eq{eqn:unm}.

\subsection{Pole-shift: diagonal dRSE}
\label{1stOrder}

To treat a shift in the pole positions of the permittivity, the idRSE requires the shifted pRSs to be included in the basis, along with the
RSs of the unshifted poles. This can add a significant numerical effort, as the matrix diagonalization required to solve \Eq{eqn:RSE} has the computational complexity ${\cal O}(N^3)$. To avoid this, one can use a diagonal version of the dRSE, only retaining the diagonal elements in \Eq{eqn:RSE}. We call this method ddRSE. It is linked to the first-order perturbation theory \cite{DoostPRA14, WeissPRL16}, which is also neglecting all off-diagonal elements. Accordingly, we expect this approach to be suited to treat small shifts in resonance frequency of the permittivity, such as the change of optical phonon frequencies with pressure or temperature. Neglecting the off-diagonal elements of \Eq{eqn:RSE} yields
\be
\omega = \omega_n \frac{1-U_{nn}}{1+V_{nn}}
\ee
Considering only the resonance shift of one pole $\Omega_1$, without changing conductivities or $\varepsilon_\infty$, all $V_{nm}$ vanish, and
\be
U_{nm} \!=\! \sum_{j=\pm1}\!\left[ \frac{i}{\omega_n-\widetilde{\Omega}_j} - \frac{i}{\omega_n-\Omega_j} \right] \!\int \!\En(\br)\cdot \hsigma_j(\br)\Em(\br)d \br\,,
\label{Unm-shift}
\ee
where $\widetilde{\Omega}_1$ is the perturbed pole position.

\begin{figure}
\centering
\includegraphics[width=\columnwidth]{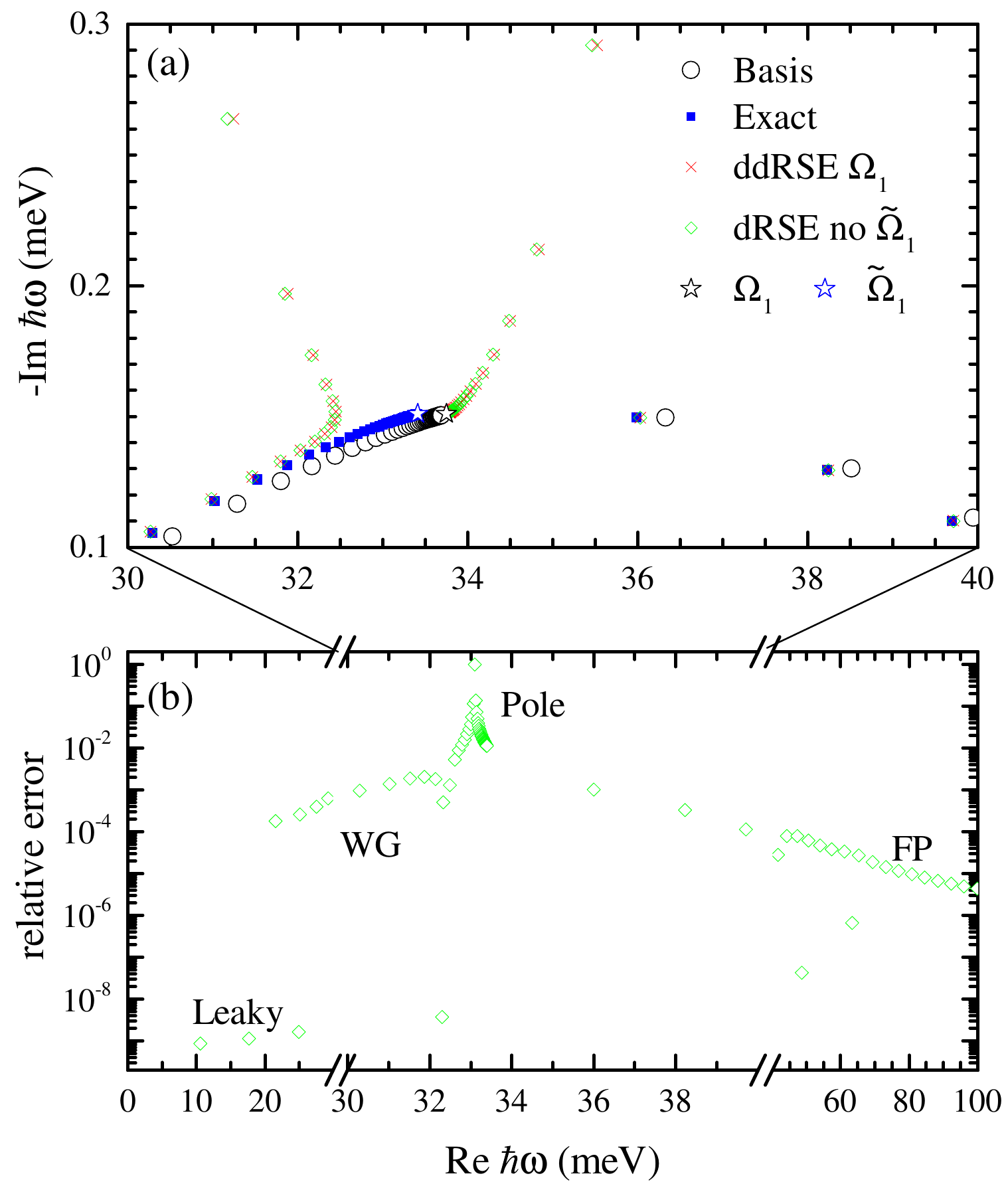}
\caption{(a) Pole-related RS frequencies of the phonon Lorentz pole of the permittivity describing the optical-phonon range in GaAs. The parameters of the GaAs microsphere and of the RSs shown are the same as in \Fig{fig:GaAs50micron}, with the pole of the permittivity shifting by 1\% of its real part from $\Omega_1$ (black star) to $\widetilde{\Omega}_1$ (blue star). Dispersive basis RSs (circles), exact RSs (squares), ddRSE (crosses) and  dRSE not including the infinitesimal-dispersive pRSs at $\widetilde{\Omega}_1$ in the basis (diamonds), with $N = 315$ corresponding to $\Omega_c = 0.4$\,eV. (b) Relative error for the RSs calculated using the ddRSE.
}
\label{fig:poleShift}
\end{figure}

To evaluate the suitability of this approach, we use the
optical phonon-induced dispersion in GaAs discussed in \Sec{GaAs}.
We consider a shift of the phonon frequency by $-1$\%, which is applied to the real part of $\Omega_1$, in order to obtain $\widetilde{\Omega}_1$. We note that the GaAs phonon frequency under isotropic pressure shifts by 0.49\,meV/GPa \cite{AdachiBook94}, so that a 1\% change would corresponds to exerting about 0.7\,GPa pressure.

The resulting RSs close to $\Omega_1$ are shown in \Fig{fig:poleShift}(a). The exact perturbed RSs converge towards the shifted pole $\widetilde{\Omega}_1$, as expected. The diagonal (first order) approach is reproducing the exact RSs well when they are sufficiently far from the pole. Close to the pole instead, large errors are encountered. Notably, using all off-diagonal elements defined by \Eq{Unm-shift} and solving the eigenvalue problem \Eq{eqn:RSE} does not significantly reduce the errors. This indicates that the effect of the basis incompleteness, missing the pRSs of $\widetilde{\Omega}_1$, is dominating the remaining error.

The relative error of the ddRSE results compared to the exact solutions is shown in \Fig{fig:poleShift}(b). We find that the error for the WG modes is decreasing with increasing separation from $\Omega_1$. The leaky modes have a small error since they have little spatial overlap with the sphere, and are spectrally broad, making them insensitive to a shift of $\Omega_1$. We can thus conclude that the diagonal approximation, equivalent to a first-order perturbation theory, is a valid approach for small pole shifts, and separations of the RSs of interest from the poles much larger than these shifts.

The physical properties in scattering and absorption are expected to be dominated by the RSs away from the poles of the permittivity, as close to the poles the permittivity is very high, screening the outside field. Furthermore, for the response at real frequencies, the specific position of the RSs close to the poles is having a minor influence, since the poles typically have a significant imaginary part, and are therefore separated from the real axis.

\subsection{Dispersive RSE: Absorption to gain}
\label{Gain}

Another interesting feature one can address using the dRSE is the transition from absorption to gain, leading to lasing.  In terms of the RSs, lasing occurs once the imaginary part of the RS frequency becomes positive, indicating an exponential increase of the field with time. This increase of the field comes with a corresponding increase of stimulated transitions, which in realistic systems results in gain saturation, reducing the imaginary part of the RS frequency to zero in a steady-state state regime of lasing \cite{TureciS08,ChowBook12}.

In the classical model of Lorentz oscillators, poles representing absorption have positive imaginary conductivity~\cite{SehmiPRB17}. Continuously changing this conductivity to  negative imaginary values transforms the absorption to gain. When representing the absorption due to interband transitions in a solid, a prefactor $1-\fe-\fh$ is applied to the conductivity, where $\fe$ and $\fh$ are the occupation functions of, respectively, the related electron state in the conduction band and hole state in the valence band, which in thermal equilibrium are given by the Fermi distribution function \cite{ChowBook12}. Creating the population inversion changes the sign of the conductivity by changing the occupation factor $1-\fe-\fh$ from 1 for no occupation ($\fe=\fh=0$) to $-1$ for full occupation ($\fe=\fh=1$).

\begin{figure}
\includegraphics[width=\columnwidth]{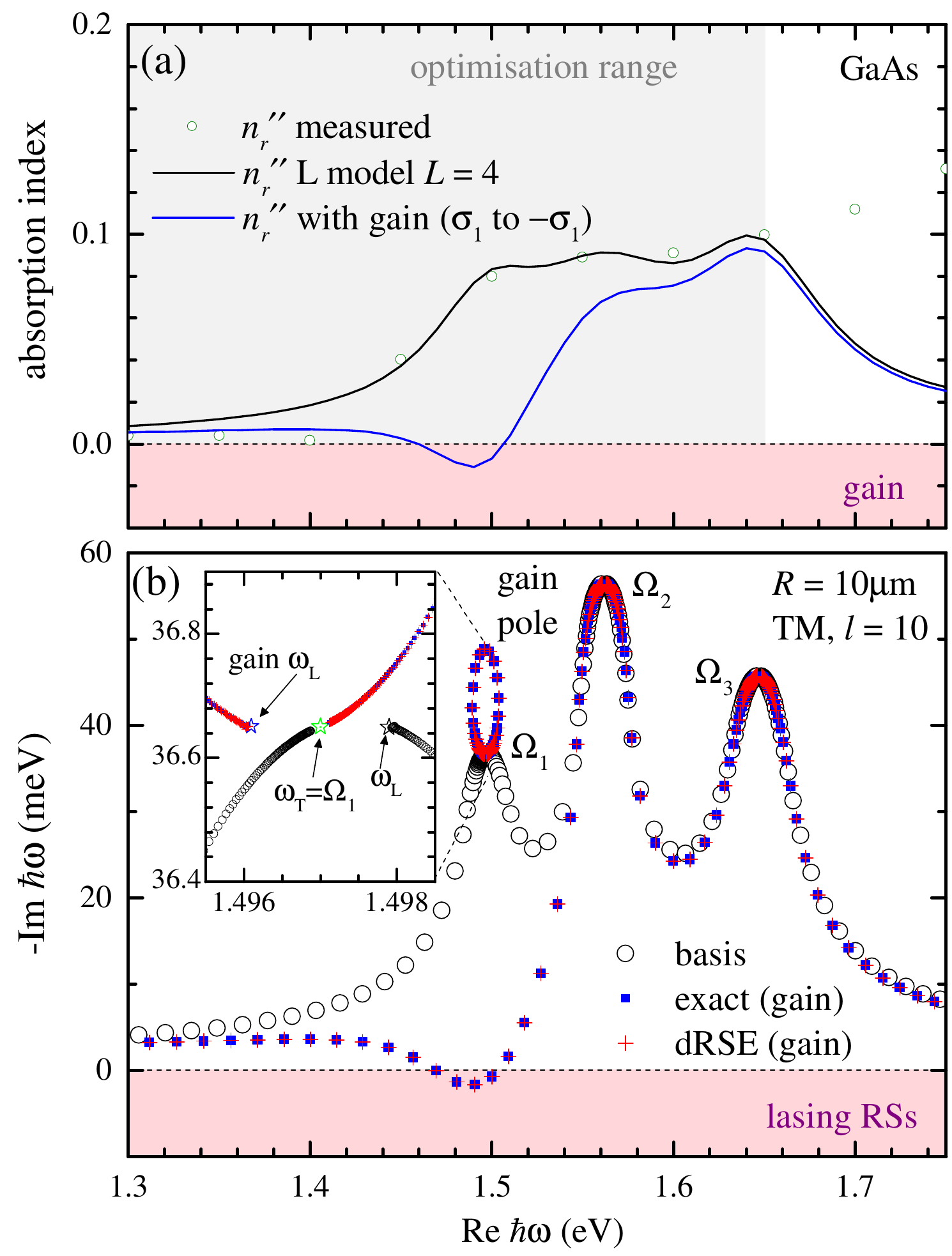}
\caption{(a) Absorption index $n_r''$
of GaAs around the band gap~\cite{Sopra} (circles) and its fit with the Lorentz model \Eq{eqn:L} for $L=4$ (red line) fitted in the range 1.3--1.65\,eV, as functions of the photon energy $\hbar\omega$. Inverting the conductivity of the pole at $\Omega_1$ (blue line) exhibits gain seen as negative absorption index around 1.49\,eV. (b)  RSs for absorption and gain, as labelled: exact RSs at absorption, used as basis for the dRSE (open circles), exact RSs at gain (blue squares), and dRSE RSs at gain (red crosses), for a basis size of $N= 2496$, corresponding to $\Omega_c = 7.89$\,eV. The inset shows the RSs close to the pole $\Omega_1$, and the related longitudinal optical frequencies $\omega_{\rm L}$ at which $n_r$ vanishes, for absorption and gain.}
\label{fig:GaAs10micron}
\end{figure}

As an example, we use GaAs, which is a direct band-gap semiconductor widely used in semiconductor lasers. To model the transition from absorption to gain, we concentrate on the energy region close to the band gap, where a population inversion is created by optical or electrical pumping, and fit the permittivity data~\cite{Sopra} in the range 1.3 to 1.65\,eV with four pairs of Lorentz poles having purely imaginary conductivities. The measured absorption index $n_r''$ and the resulting fit of the Lorentz model are shown in \Fig{fig:GaAs10micron}(a). We see that the fit is not as good as for the optical-phonon range shown in \Fig{fig:GaAs50micron}(a), and allowing the conductivities to have non-zero real parts could lead to a better agreement, as in the case of gold. However, we intentionally fit the permittivity with a model of classical Lorentz oscillators, in order to enable a simulation of the population inversion. The fit function contains three poles close to the band-gap having a small linewidth to describe the sharp increase of the absorption at the band-edge, and the fourth pole at higher energy outside the fit range, providing a background. When changing the sign of the first-pole conductivity $\sigma_1$, in order to simulate the change from no to full occupation, the resulting absorption index shows a negative region around 1.49\,eV, indicating the region where the material exhibits gain.

\begin{figure}
\includegraphics[width=\columnwidth]{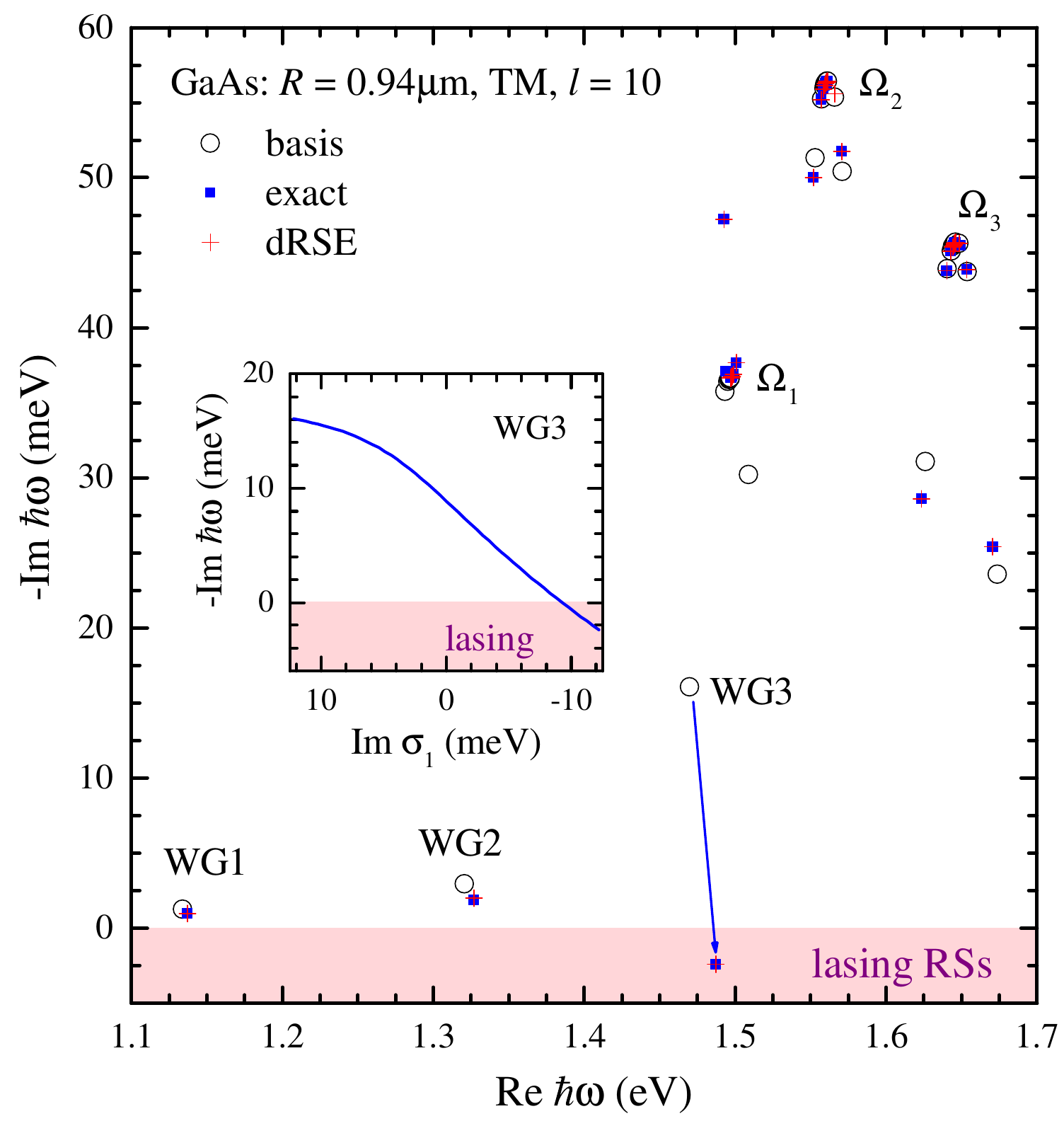}
	\caption{(a) Absorption to gain RSE for GaAs showing the evolution of a RS (WG3) from negative to positive values of $\text{Im}\,\hbar\omega$ as well as other nearby modes. The inset shows the evolution of the imaginary part of WG3 versus the $\Omega_1$-pole conductivity $\text{Im}\,\sigma_1$.}
	\label{fig:GaAs940nm}
\end{figure}

We first consider a rather large microsphere of radius $R=10\,\mu$m, and choose $l=10$. The wavelength in the medium resonant to the gain  is about $\lambda=0.24\,\mu$m, much smaller than the radius, and thus the RSs close to the gain pole are of FP type for the chosen $l$.
To provide a round-trip amplification, the losses due to the transmission across the sphere surface must be compensated by the gain through the sphere. The amplitude reflection coefficient for normal incidence is $r=(n_r'-1)/(n_r'+1)\approx 0.55$, using the refractive index of $n_r'=3.5$. The amplification of the wave amplitude propagating one diameter is $g=\exp(-2R n_r''\omega/c)$. The condition for round-trip amplification is then $gr>1$, which requires $n_r''<-0.004$. We see in \Fig{fig:GaAs10micron}(a) that this condition is fulfilled around 1.49\,eV.

The RSs under this gain, calculated using the dRSE with the dispersive basis of RSs in the absorption regime (using a basis size $N = 676$ corresponding to $\Omega_c = 26.78$\,eV), are shown in  \Fig{fig:GaAs10micron}(b) over the same energy range as used in \Fig{fig:GaAs10micron}(a). Away from the poles, the basis shows nearly equidistant FP modes, which have an imaginary part following the losses due to the absorption index $n_r''$. In the absorption regime, close to the poles, the RSs converge from the lower frequency side towards the poles, which are at the transversal optical resonance frequencies $\omega_{\rm T}$,  and show gaps at higher frequencies up to the longitudinal optical frequency $\omega_{\rm L}$, where $n_r=0$, similar to the case of the optical phonons in \Fig{fig:GaAs50micron}(b). Note that in the present case, the set of RSs approaching the longitudinal optical frequencies is rather large, in the order of 100. This is due to the large radial quantum number of the basis FP modes in the region of the pole frequency. This number can be estimated as $4R/\lambda\approx 166$, using $\lambda=0.24\,\mu$m.

After switching to gain, some RSs in the gain spectral region show a positive imaginary part, as expected from the above discussion, and are thus lasing RSs. Furthermore, a curious arrangement of RSs is observed close to the gain pole, which occurs due to the inverted influence of the pole -- the refractive index now increases approaching the pole from higher frequencies, so that the infinite series of RSs now converges towards the pole from the higher frequency side. These RSs, however, form a loop, moving clockwise over the pole, starting at a value around the corresponding longitudinal optical frequency $\omega_{\rm L}$, which is now on the lower frequency side of the pole, as shown in the inset of \Fig{fig:GaAs10micron}(b). The original, unperturbed series of FP modes, present without the pole, is instead repelled from the pole.
This is qualitatively different from the case of absorption, where the RS series is attracted to the pole, splitting the RSs into an infinite set of all radial quantum numbers below the pole frequency $\omega_{\rm T}=\Omega_1$, and a second infinite set of all radial quantum numbers above $\omega_{\rm L}$.

To investigate the regime of WG mode lasing at $l=10$, we consider in \Fig{fig:GaAs940nm} a much smaller sphere of radius $R=0.94\,\mu$m. In this case, the gain region lines up with a WG mode in the basis (WG3), moving it to a positive imaginary energy. The evolution of the imaginary part of WG3 with the change from absorption to gain is shown in the inset. We observe that with the build-up of the gain, given by the decreasing imaginary part of the conductivity, the imaginary part of the WG3 energy moves continuously towards positive values, eventually reaching the lasing regime. The other RSs only change slightly since they are not close to $\Omega_1$, so that WG3 remains the only lasing mode of the system.

\section{Conclusions}

We have shown a set of applications of the dispersive RSE, a method developed for accurate and efficient calculation of the resonant states of optical systems exhibiting frequency dispersion. We have also presented the infinitesimal-dispersive RSE, which uses the resonant states of a non-dispersive system as a basis for treating an optical system with dispersion, adding new poles in the frequency dispersion in the limit of zero conductivity. This requires an extension of the RSE basis, in order to include the so-called pole resonant states which are all at the pole frequency.

We apply the method to calculate the resonant states of spheres made of materials described by a frequency-dependent permittivity in the form of a generalized Drude-Lorentz model.
Using our fit program~\cite{SehmiPRB17} accurately determining the parameters of the model, we fit gold with one Drude and three pairs of Lorentz poles. Using the infinitesimal-dispersive RSE, we perturb a 10\,nm-radius nanosphere of sand into gold, with the error in the resonant-state frequencies scaling as the inverse cube of the basis size. We also analyse the evolution of the surface plasmon polaritons with the size of the gold nanosphere, showing the effects of radiation and retardation with increasing size.

We accurately describe the optical phonon range in GaAs with a single pair of Lorentz poles and perturb a 50\,$\mu$m-radius dielectric sphere to GaAs showing the reststrahlen band near the pole and a series of resonant states approaching the pole. Fitting the band gap range 1.3--1.65\,eV of GaAs with four pairs of classical Lorentz model poles, we perturb the lowest energy pole from absorption to gain and identify lasing resonant states.
In general, each Lorentz pole of the permittivity creates an additional infinite series of resonant states with the full set of radial quantum numbers corresponding to frequencies starting from a value around the longitudinal frequency at which the refractive index vanishes and extending towards infinity or converging to the next Lorentz pole at which the index diverges. Adding an absorption pole attracts the non-dispersive series, splitting it into two infinite series, above and below the pole. Adding a gain pole instead repels the non-dispersive series, and creates an additional series forming a clockwise loop in complex frequency, starting from the longitudinal frequency left of the pole and converging towards the pole from the right.

Finally, we introduce a diagonal dispersive RSE, a first-order perturbation method for treating small changes in the pole position, which is suited for the resonant states sufficiently separated from the pole of the dispersion and which has a much lower numerical complexity than the full dispersive RSE.

We believe that the dispersive RSE, once implemented in a accessible software package, including arbitrary 3D open systems \cite{DoostPRA14}, and scattering and absorption with external excitation \cite{LobanovPRA18}, will be a powerful and widely used tool for treating electromagnetic problems.

\acknowledgments

This work was supported by the S\^er Cymru National Research Network in Advanced Engineering and Materials and by the EPSRC under grant EP/M020479/1. E.A.M. acknowledges support by RBRF (Grant No. 16-29-03282).

\appendix

\section{Drude-Lorentz model of the permittivity and fit for Au and GaAs}
\label{App:DL}

Implementing causality, the dispersive permittivity obeys the relation $\heps^\ast(\br,\omega)=\heps(\br,-\omega^\ast)
$. This implies that poles appear in pairs in the permittivity \Eq{eqn:eps}. In particular, each pole at $\Omega_j$ has a partner at $\Omega_{-j}=-\Omega^\ast_j$ with $\hsigma_{-j}=\hsigma^\ast_j$. A pole with Re\,$\Omega_j=0$, such as that of the Drude model, can be treated as a pair of poles merged to produce a single pole on the imaginary $\omega$-axis with a purely real conductivity. Furthermore, Lorentz reciprocity requires that all tensors in \Eq{eqn:eps} are symmetric.
%~\Cite{LandauLifshitzV8Book84}.
For isotropic materials, such as gold and GaAs, these tensors can be replaced by scalars.

In the calculations presented in this paper, we use for GaAs a Lorentz model of the permittivity consisting of $L$ pairs of poles,
%which consists of a single Drude pole and and multiple Lorentz poles, written as \cite{SehmiPRB17}
\be
\varepsilon(\omega)=\varepsilon_\infty+\sum_{j=1}^L\left(\frac{i\sigma_j}{\omega-\Omega_j} + \frac{i\sigma_j^\ast}{\omega+\Omega_j^\ast}\right)\,,
\label{eqn:L}
\ee
and for gold a Drude-Lorentz model,
\be
\varepsilon(\omega)=\varepsilon_\infty- \frac{\gamma\sigma}{\omega(\omega+ i\gamma)}  +\sum_{j=1}^L\left(\frac{i\sigma_j}{\omega-\Omega_j} + \frac{i\sigma_j^\ast}{\omega+\Omega_j^\ast}\right)\,,
\label{eqn:DL}
\ee
in which the Drude term of the dispersion has been added, consisting of Ohm's law pole at $\omega=0$ and a Drude pole at $\omega=\Omega_0=-i\gamma$:
\be
\frac{i\sigma}{\omega} - \frac{i\sigma}{\omega+ i\gamma}=- \frac{\gamma\sigma}{\omega(\omega+ i\gamma)}\,.
\ee
Here $\varepsilon_\infty$ is the permittivity at high frequencies and $\sigma$ is the real DC conductivity. The generalized conductivities $\sigma_j = \sigma'_j +i\sigma''_j $ at the Lorentz poles $\Omega_j=\Omega_j'+i\Omega_j''$  are complex.
% We denote real and imaginary parts of complex quantities with prime and double prime, respectively, and keep using this notation throughout the paper.

\Tab{tab:para4} contains the parameters of the fits shown in Figs.\,\ref{fig:gold10nm}, \ref{fig:GaAs50micron}, and \ref{fig:GaAs10micron}, using the Drude-Lorentz and the Lorentz models with $L=3$, 1, and 4, respectively. The resulting root-mean square error $S$ of the fit is also shown. Its definition, with and without experimental errors taken into account in the optimization, is provided in \Onlinecite{SehmiPRB17}. This difference in the definition leads, in particular, to a much higher value of $S$ for gold, in which case the fit takes the measurement errors~\cite{JohnsonPRB72} into account.

\begin{table}
	\centering
	\begin{tabular}{>{$}c<{$} | >{$}c<{$} | >{$}c<{$} | >{$}c<{$}}
		\text{Material}				& \text{Au}       & \text{GaAs*}		& \text{GaAs**}\\
		\hline
		\rule{0pt}{3ex}
		\varepsilon_\infty	& 0.5		& 11.0	& 8.6013\\
		\gamma\ ({\rm eV})	& -0.065748	& -		& -\\
		\sigma\ ({\rm eV})	& 1133.0	& -		& -\\
		\hline
		\rule{0pt}{3ex}
		\Omega_1'\ ({\rm eV})	& 2.5936	& 0.033314	& 1.497\\
		\Omega_1''\ ({\rm eV})	& -0.41875	& 1.4904 \times 10^{-4}	& -0.03665\\
		\sigma_1'\ ({\rm eV})	& 1.4029	& 0.0		& 0.0\\
		\sigma_1''\ ({\rm eV})	& 0.76857	& 0.033262	& 0.01224\\
		\hline
		\rule{0pt}{3ex}
		\Omega_2'\ ({\rm eV})	& 3.8192	& -		& 1.5612\\
		\Omega_2''\ ({\rm eV})	& -1.3246	& -		& -0.05643\\
		\sigma_2'\ ({\rm eV})	& 0.41939	& -		& 0.0\\
		\sigma_2''\ ({\rm eV})   & 4.5468	& -		& 0.02432\\
		\hline
		\rule{0pt}{3ex}
		\Omega_3'\ ({\rm eV})	& 9.6899	& -		& 1.6463\\
		\Omega_3''\ ({\rm eV})	& -4.2933	& -		& -0.0457\\
		\sigma_3'\ ({\rm eV})	& 0.012244	& -		& 0.0\\
		\sigma_3''\ ({\rm eV})	& 14.817	& -		& 0.02404\\
		\hline
		\rule{0pt}{3ex}
		\Omega_4'\ ({\rm eV})	& -		& -		& 2.2853\\
		\Omega_4''\ ({\rm eV})	& -		& -		& -0.00778\\
		\sigma_4'\ ({\rm eV})	& -		& -		& 0.0\\
		\sigma_4''\ ({\rm eV})	& -		& -		& 2.9302\\
		\hline
		\rule{0pt}{3ex}
		\hbar\omega_1\ ({\rm eV})	& 0.64	& 0.031	&1.3\\
		\hbar\omega_N\ ({\rm eV})	& 6.6		& 0.0372	&1.65\\
		N_d					& 49		& 47		&8\\
		S					& 1.4795	& 0.0372 	&0.0055\\
	\end{tabular}
	\caption{Optimized model parameters for Au and GaAs phonon (*) and band-edge (**) ranges, using the fit function with optimization energy ranges corresponding to the data shown in Figs.\,\ref{fig:gold10nm}, \ref{fig:GaAs50micron}, \ref{fig:GaAs10micron}, and \ref{fig:GaAs940nm}. The number of data values $N_d$ and the resulting error $S$ are also given.}
	\label{tab:para4}
\end{table}

\section{TM modes of a sphere, their normalization and matrix elements}
%\section{RSs of a dielectric sphere and infinitesimal-dispersive basis} % metal sphere?
\label{App:TMmodes}

The electric field of a TM mode of a sphere of radius $R$ in vacuum has the form (in spherical coordinates)
\be
\mathbf{E}(\br)=\dfrac{A_{nl}}{n_r k_n r}\left(
\begin{array}{ccc}
l(l+1)\,\psi_{nl}(r)Y_{lm}(\theta,\varphi)\\[5pt]
\dfrac{\partial}{\partial r} r \,\psi_{nl}(r)\dfrac{\partial}{\partial\theta}Y_{lm}(\theta,\varphi)\\%[10pt]
\dfrac{\partial}{\partial r} r \,\psi_{nl}(r)\dfrac{1}{\sin\theta}\dfrac{\partial}{\partial\varphi}Y_{lm}(\theta,\varphi)\\
\end{array}
\right), \label{eqn:E_TM}
\ee
where $Y_{lm}(\theta,\varphi)$ are real-valued spherical harmonics~\cite{DoostPRA14}, and the radial wave functions within the sphere, $r\leqslant R$, are given by
\be
\psi_{nl}(r)=\frac{j_l(n_r k_n r)}{j_l(n_r k_n R)}\,.
\ee
Here $k_n=\omega_n/c$ is the RS wave number and $n_r$ is the refractive index, which is given by
\be
n_r^2=\varepsilon(\omega_n)\,,
\label{nr}
\ee
with the permittivity $\varepsilon(\omega)$.
The normalization constant $A_{nl}$ is found from the general RS normalization condition with dispersion, \Eq{norm}, and has the form~\cite{MuljarovPRB16a}
\be
\frac{1}{A^2_{nl}}=l(l+1)R^2(n_r^2-1)n_r^2 D_{nl}
\label{A-normTM}
\ee
with
\be
D_{nl}=\frac{1}{n_r^2}\left[\frac{j_{l-1}(x)}{j_l(x)}-\frac{l}{x}\right]^2+\frac{l(l+1)}{x^2}+\eta_n C_{nl}
\ee
and
\be
(n_r^2-1)C_{nl}=-\frac{2l}{x^2}+\frac{j_{l-1}^2(x)}{j_l^2(x)}-\frac{j_{l-2}(x)}{j_l(x)}\,,
\ee
where $j_l(x)$ are the spherical Bessel functions, $x=n_r\omega_nR/c$,
\be
\eta_n=\left.\frac{\omega}{2\varepsilon(\omega)}\frac{\partial\varepsilon(\omega)}{\partial \omega}\right|_{\omega=\omega_n}\,,
\label{eta}
\ee
and $n_r$ is given by \Eq{nr}.

The RS eigenfrequencies $\omega_n$ are solutions of the secular equation for the TM polarization
\be
\frac{1}{n_r}\frac{j_{l-1}(n_r z)}{j_l(n_r z)}=\frac{h_{l-1}(z)}{h_l(z)}-\frac{l}{z}\left(1-\frac{1}{n_r^2}\right),
\label{secular}
\ee
in which $h_l(z)\equiv h^{(1)}_l(z)$ is the spherical Hankel function of the first kind, with $z=\omega_n R/c$, and $n_r(\omega_n)$ is defined by \Eq{nr}.

For small values of $R$, corresponding to the electrostatic limit, the spherical Bessel and Hankel functions can be expanded into power series of their arguments and keeping only the leading terms, which is valid for $|\omega| R/c\ll1$ and $|n_r(\omega)\omega| R/c\ll1$, one can obtain from \Eq{secular} a well-known equation
\be
\varepsilon(\omega)=-\frac{l+1}{l}\,,
\label{equ:SP}
\ee
determining for each $l$ the eigenfrequency $\omega$ of the surface plasmon in the low-frequency (electrostatic) limit.

For the pRSs, due to a pole of the dispersion at $\Omega_j$, all the eigenfrequencies are the same, $\omega_n=\Omega_j$, but the eigenvalues of the refractive index $n_r(\omega_n)$ are different. They are again determined by the secular equation (\ref{secular}) which is now solved for $n_r$ while keeping $z=\Omega_j R/c$ fixed.
These eigenvalues $n_r$ are linked to $q_n$, as defined by \Eq{qn},
\be
n^2_r(\omega_n)=\varepsilon_\infty+\sum_{j'\neq j}\frac{i\sigma_{j'}}{\Omega_j-\Omega_{j'}} +\frac{1}{q_{n}}\,,
\label{qn1}
\ee
and to the values $\eta_n$ in the normalization constants \Eq{eta}:
\be
\eta_n=-\frac{1}{\xi}\,\frac{\Omega_j}{2 n_r^2 q^2_n}=-\frac{1}{\alpha_n^2}\,\frac{1}{2 n_r^2 q_n }\,,
\ee
in accordance with Eqs.\,(\ref{eps-der}) and (\ref{eqn:alpha}), using $\xi=\alpha_n^2\Omega_j/q_n$. Clearly, $\xi\to0$ makes the normalization constants $A_{nl}$ vanishing.
Re-defining these  constants as
\be
A_{nl}=\alpha_n \tilde{A}_{nl}\,,
\label{Anlt}
\ee
in accordance with \Eq{Ent}, we find the new (finite) normalization constants of the pRSs:
\be
\frac{1}{\tilde{A}^2_{nl}}=-{l(l+1)R^3 (n_r^2-1)}\frac{1}{2q_n}\,C_{nl}\,.
\label{Apole}
\ee

For any homogeneous perturbation of the sphere, all the matrix elements, contributing to \Eq{eqn:RSE} or \Eq{eqn:epsDispRSE} are given by the integrals
\be
W_{nm} = \int_{V_R} \tEn(\br)\cdot\tEm(\br)\,d \br\,,
\ee
where $V_R$ is the volume of the sphere. These integrals
have the following analytic form:
\be
W_{nm}=\tilde{A}_{nl}\tilde{A}_{ml} l(l+1) R^3 F_l(x,y)\,,
\label{Wnm}
\ee
where $x=n_r(\omega_n) \omega_n R/c$ and $y=n_r(\omega_m) \omega_m R/c$, and both RSs $n$ and $m$ belong to the same orbital quantum number $l$. The function $F_l(x,y)$ is defined as
\be
F_l(x,y)=\frac{1}{x^2-y^2}\left[x\frac{j_{l-1}(y)}{j_l(y)}-y\frac{j_{l-1}(x)}{j_l(x)}\right] -\frac{l}{xy}
\ee
which for coinciding arguments becomes
\be
F_l(x,x)=\frac{1}{2}\left[\frac{j^2_{l-1}(x)}{j^2_l(x)}-\frac{j_{l-2}(x)}{j_l(x)}\right] -\frac{l}{x^2}\,.
\ee
Note that using the normalization coefficients \Eq{Apole} for the pRSs, the diagonal elements \Eq{Wnm} become $W_{nn}=-q_n$, in accordance with their normalization condition
\be
1=-\frac{1}{q_n}\int_{V_R}\tEn^2 (\br)\,d{\bf r}\,,
\ee
see \Eq{polenorm}.

\section{Calculating the basis RSs}
%\section{RSs of a dielectric sphere and infinitesimal-dispersive basis} % metal sphere?
\label{App:basis}

To generate a basis of unperturbed RSs, the secular equation (\ref{secular}) is solved numerically using the Newton-Raphson algorithm. This relies on a suited choice of guess values, which for $|n_r z|\lesssim 1$ are found on a two-dimensional grid in the complex $z$ plane. For large values of $n_r z$, however, we use approximate recursive relations given below, which provide reliable guess values based on the eigenvalues found for the neighboring RSs. To derive these relations, it is convenient to modify \Eq{secular} to the following form:
\be
\frac{1}{n_r}\frac{j'_{l}(n_r z)}{j_l(n_r z)}=\frac{h'_{l}(z)}{h_l(z)}+\frac{1}{z}\left(1-\frac{1}{n_r^2}\right),
\label{secular1}
\ee
where $j'_{l}(x)$ and $h'_{l}(x)$ are the derivatives of the spherical Bessel and Hankel functions with respect to their arguments. Noting that
\be
j_{l}(x)\approx \frac{1}{x}\cos\left(x-\frac{l+1}{2}\pi\right)
\ee
at large $x$, we find
\be
\frac{j'_{l}(x)}{j_{l}(x)}\approx -\tan\left(x-\frac{l+1}{2}\pi\right),
\ee
which provides an estimate for the left-hand side of \Eq{secular1}. Its right-hand side is in turn a slowly changing function. Therefore, the neighboring RSs are separated by $\Delta x\approx \pi$, which is the period of the tangent function. This separation can be further evaluated as
\be
\Delta x=n_r(z+\Delta z)(z+\Delta z)-n_r(z)z \approx (n_r+n_r' z)\Delta z\,,
\ee
where $n_r'=dn_r/dz$. Then we find approximate relations for the neighboring roots of \Eq{secular} in the following three cases:

\noindent
(i) No dispersion of permittivity or RS eigenfrequencies are far from any poles of the dispersion, i.e. $n_r$ is nearly constant:
\be
\Delta z\approx \frac{\pi}{n_r}\,.
\label{deltaz}
\ee

\noindent
(ii) Strong dispersion, such as that for the RS near the pole of the dispersion:
\be
\Delta z\approx \frac{\pi}{n_r+n_r' z}\,.
\ee

\noindent
(iii) Pole RSs -- which are required for constructing an infinitesimal-dispersive basis. In this case \Eq{secular} is solved for $n_r$ with $z$ fixed. For the neighboring pRSs, we find
\be
\Delta n_r\approx \frac{\pi}{z}\,.
\label{deltan}
\ee

\end{document}